\documentclass[cite]{epl}

\usepackage{graphicx}
\usepackage{dcolumn}
\usepackage{bm}
\usepackage{rotating} 

\usepackage{amssymb}
\usepackage[latin1]{inputenc}

\title{Pressure-Induced Simultaneous Metal-Insulator and Structural-Phase 
Transitions in LiH: a Quasiparticle Study}
\shorttitle{Phase transition properties of lithium hydride}
\author{S.~Leb\`egue\inst{1,4} \and B.~Arnaud\inst{2} \and M.~Alouani\inst{1} \and W.~E.~Pickett\inst{4}}
\author{S.~Leb\`egue\inst{1,4} \and M.~Alouani\inst{1} \and B.~Arnaud\inst{2} \and W.~E.~Pickett\inst{4}}
\shortauthor{S.~Leb\`egue \etal}
\institute{
\inst{1} Institut de Physique et de Chimie des Mat\'eriaux de
Strasbourg (IPCMS), UMR 7504 du CNRS, 23 rue du Loess, 67037 Strasbourg, France, EU\\
\inst{2} Groupe Mati\`ere condens\'ee et Mat\'eriaux
(GMCM), Campus de Beaulieu - Bat 11A
35042 Rennes cedex, France, EU\\
\inst{4} Department of Physics, University of California, Davis, CA 95616, USA }
\begin{document}
\maketitle
\begin{abstract}
A pressure-induced simultaneous metal-insulator transition  (MIT) and structural-phase 
transformation  in 
lithium hydride with about  1\% volume collapse  has been predicted  by means of 
the local density approximation (LDA) 
in conjunction with an all-electron  GW approximation method.  
The LDA wrongly predicts that the MIT
 occurs before the structural phase transition.   As a byproduct, 
it is shown that only the use of the generalized-gradient approximation together with
the zero-point vibration produces an equilibrium lattice parameter, bulk modulus, and 
an equation of state  that are in excellent agreement with experimental results. 
\end{abstract}
Lithium hydride is probably the simplest compound that exists:
a strongly ionic crystal  with four electrons per unit cell and  
crystallizing in the rocksalt structure, the so-called B1 phase.  
Despite this simplicity, LiH and its isotopes are 
attractive for the study of solid state properties, e.g., electronic structure, 
lattice vibration, and defect properties. In addition, 
possible technological applications
have motivated extensive studies in the past, as reviewed by Islam\cite{review}.
In particular, the metal-insulator transition (MIT) has 
been studied by several 
groups\cite{review,griggs,behringer,olinger,hama,hammerberg,kondo,hochheimer}.
In few alkali hydrides a structural phase transition (PT) from the B1 phase to the 
B2 (CsCl structure) phase  was determined experimentally\cite{review,hochheimer} 
and theoretically\cite{hama,rodriguez,ahuja} within the 
local-density approximation.
It is only recently that LiH was found to exhibit the same type of PT
 but at a much higher pressure\cite{ahuja}. Despite this extensive study we 
believe, as it will be shown later, that this  PT
is not well understood. 
The purpose of this Letter  is then to present results improving the
current understanding of the electronic structure and the MIT 
in alkali hydrides, using LiH as a prototype. 
In particular, we address the issue of  the pressure-induced MIT 
 by gap closure and structural PT\cite{thermodynamics}, and  
investigate the different levels of approximations to the total energy,   
aiming to predict correctly the equilibrium lattice parameter, the bulk modulus, 
and the  equation of state (EoS) of LiH.

The MITs were classified by Chacham, Zu, and Louie\cite{czl} 
(CZL) as occurring through
one of the three following processes: (1) pressure induced structural transformation,
(2) magnetic PT of antiferromagnetic insulators, (3) gap closure 
without phase transformation. For example, xenon and hydrogen are  shown 
by the same authors, using a GW study, to exhibit a pressure-induced MIT
of type three \cite{czl}. 
One should remark, therefore, that despite the extensive studies of these kind of PTs
 within the LDA, 
they remain poorly described by this theory, 
because all these types of transitions  involve a band-gap 
closure either directly or indirectly, and it is now well established that the Kohn-Sham
density functional theory\cite{Hohenberg} drastically underestimates the band gap 
in all types of insulating materials.
In this study we will then demonstrate, using a combined total energy calculation 
and calculated excited states within the GW approximation (GWA),  
that in  LiH the MIT is caused by 
a PT from the B1  to the B2 phase (transition of type one  
according to CZL), accompanied by  a large band-gap closure and a 
volume collapse.  The LDA incorrectly predicts  an electronic 
MIT within the B1 phase, corresponding to type three PT  according 
to CZL classification, and then under further compression a structural 
transformation to the B2 phase.

Our Letter  is organized as follows.
In the first part, we investigate the structural properties of
 LiH with the all-electron Projector-Augmented-Wave method (PAW)\cite{Bloechl}.
In order to describe the  experimental ground-state properties of LiH, different parameterizations 
of the exchange-correlation functional have been used. The lattice parameter
 as well as the EoS are found to be in excellent agreement with 
 experimental data $only$ when the exchange-correlation functional is described within the
generalized-gradient approximation (GGA) and the zero-point vibration is  included 
to the total energy.  The second part of this work is devoted to the study of 
excited states using our recently implemented GW approximation\cite{lebegue}. 
In the last part  we address  the pressure-induced 
MIT issue in LiH, and show that it is  incorrectly predicted by the LDA.
\section{Electronic Ground State}
Ground-state properties of LiH have been extensively studied in the 
past\cite{review}, mainly in the local-density approximation (LDA).
In our case, we use  the PAW method, an elegant all-electron method which
keeps the simplicity of pseudopotentials but describes correctly the 
nodal region of the wave functions\cite{Bloechl}. In our calculations, 
the 1$s$ electrons of Li are also included as valence states, 
since the key role of core electrons has already been pointed out 
in Ref. \cite{bellaiche2}.
We pay particular attention to the convergence of our 
calculations\footnote{A mesh of $666$ k-points in the full Brillouin zone and an energy 
cut-off of $100$ Ry for the plane-wave basis set have been used.}. 
The exchange-correlation energy functional has been treated either in the LDA\cite{lda} 
or in the GGA\cite{gga}. 
\begin{figure}[h!]
\centerline{ \includegraphics*[angle=0,width=0.5\textwidth]{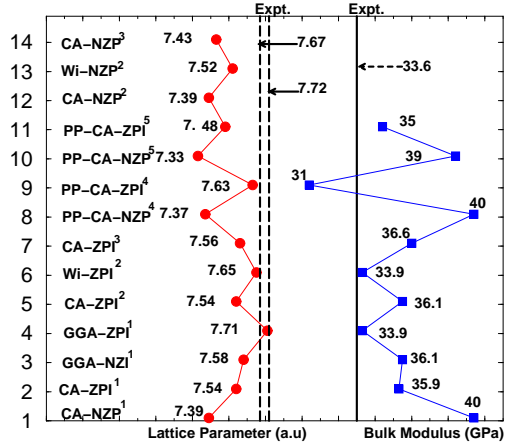}}
\caption{\label{fig:lat_parameter} 
Calculated ground-state lattice parameter $a$ (in atomic units) 
and bulk modulus (in GPa) compared with experimental\cite{exp1,exp2} and 
theoretical\cite{hama,martins,roma} results. 
The labels of the coordinates represent different types
of calculations (calculations from 1 to 7 and 12 to 14 use 
different all-electron methods and 8 and 9 pseudopotentials (PP)).
The two vertical dashed lines from left to right represent, respectively,
the experimental lattice parameters of Ref. \cite{exp1} and \cite{exp2}, and
the  vertical full line the bulk modulus of Ref. \cite{exp2}. 
The calculations neglecting or including the zero-point vibrations are 
labeled by  NZP or ZPI, respectively.  
The acronyms CA and Wi stand for the Ceperley-Adler and the Wigner parameterization 
of the LDA exchange-correlation functional, respectively. 
We obtain a good agreement with experiment $only$ when both the GGA 
and the ZPI are included in our calculations. 
}
  $^1$ Present work; $^2$ Ref. \cite{bellaiche};
  $^3$ Ref. \cite{hama};
  $^4$ Ref. \cite{martins}; 
  $^5$ Ref. \cite{roma}; 
\end{figure}
Fig. \ref{fig:lat_parameter} presents our calculated ground-state 
lattice parameter and bulk modulus compared to different 
calculations\cite{roma,bellaiche,hama,martins} and experimental
results\cite{exp1,exp2}.
Our LDA result agrees well with other calculations
performed with the same functional, as can be seen in Fig.
\ref{fig:lat_parameter}. In our case, a value
 of 7.39 atomic units (a. u.) is found, 
 but is  unsatisfactory compared  with experiment\cite{exp1,exp2}  
(about 4.3\% smaller).
This is not surprising since light atomic masses enhance the volume 
due to zero-point vibrations (ZPV) of the lattice:
\begin{equation}\label{zero_point_eq}
E_{ZP}(V) = \frac{1}{2} \sum_{i,\bf{k}} \hbar \omega_{i}({\bf k},V),
\end{equation}
where $\omega_{i}({\bf k},V)$ is the phonon frequency for a wave vector 
${\bf k}$, 
branch $i$ and  cell volume $V$, and are therefore
crucial for a proper description of the ground-state properties and the 
EoS of LiH.
In our case, we don't perform this type of calculation but instead 
extract the data from Ref. \cite{roma}.
The addition  of the zero-point vibrations (ZPV) improves
 considerably the agreement with experiment, but is still insufficient
(about $2.3 \%$ smaller).
On the other hand, the usage of the GGA as exchange-correlation functional 
improves the LDA results, but a difference of about $1.8 \%$ remains.
A combination of both the GGA and the ZPV are found to be necessary to 
obtain excellent  agreement with the experimental lattice parameter and 
bulk modulus\cite{exp1,exp2}.  
Therefore   the good agreement of   Ref. \cite{bellaiche} with experiment (
see. Fig.\ref{fig:lat_parameter})
when combining ZPV with Wigner parameterization for the LDA exchange correlation
functional is fortuitous.  This is because, as we have shown, the 
use of the GGA  highly improves  the LDA results. 
As a consequence, this agreement cannot be interpreted as a general feature
but rather, as mentioned in their work,  it is due to  a 
nearly complete cancellation of errors between the exchange and correlation
energies when the Wigner parameterization is applied to LiH.
The higher LDA value of $7.63$ a. u. reported in  Ref. \cite{martins}, which 
seems closer to the  experimental lattice parameter,  is in fact due to 
a  poor sampling in the evaluation of Eq. 
(\ref{zero_point_eq}).  Indeed, Roma {\sl et al.}  show that a good sampling 
leads  to a much lower value\cite{roma}.

\begin{figure}[h!]
\twoimages[angle=-90,width=0.5\textwidth]{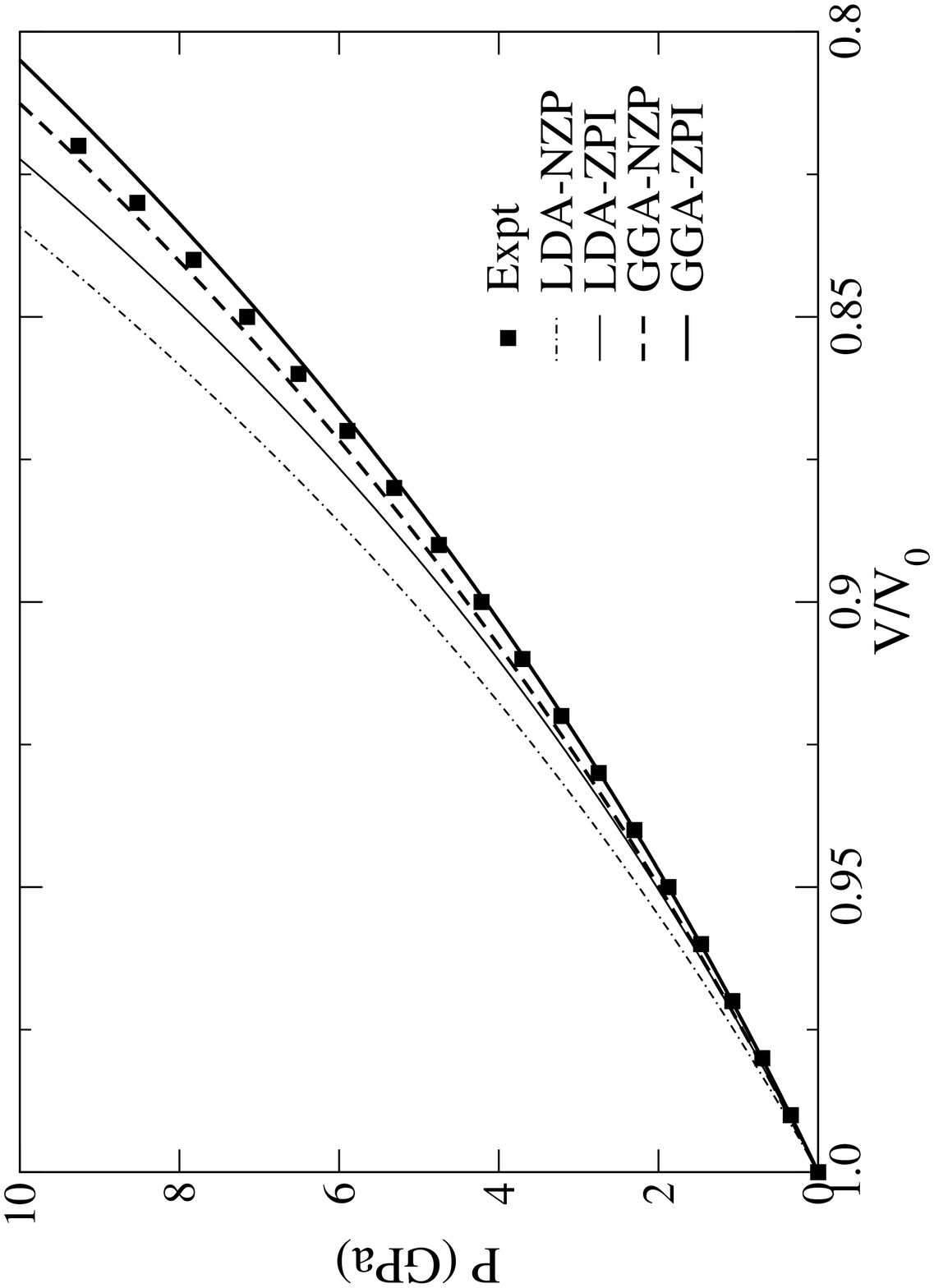}{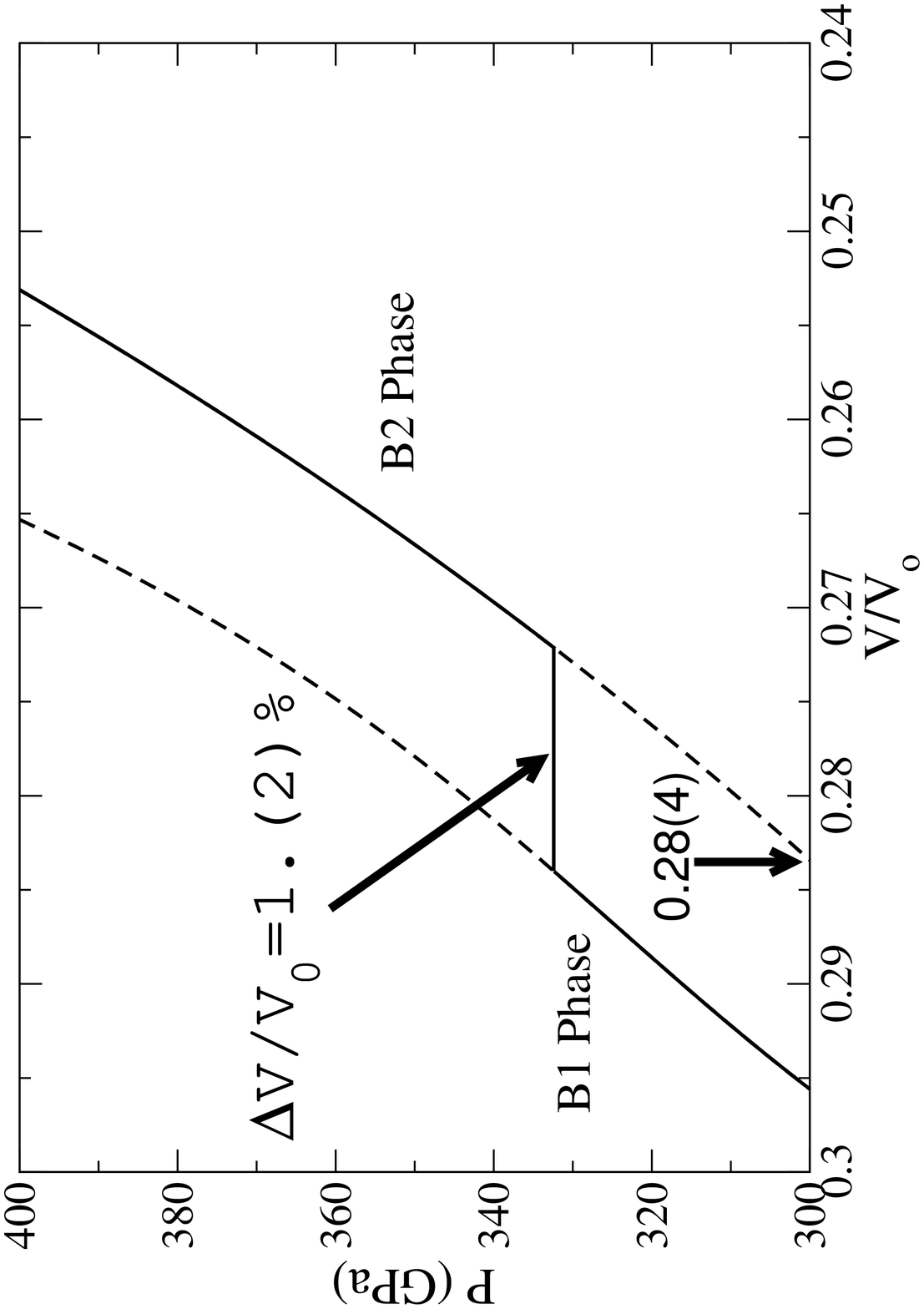}
\caption{\label{fig:eos_lih} 
The left plot shows the comparison between our calculated Birch EoS, 
using different types of approximations to the total energy, 
with experimental results (squares) of Besson\cite{besson}.
The  EoS within the GGA, including the zero-point motion, 
agrees almost perfectly with experiment in the range of 1 to 0.9 volume compression.
The right figure presents the EoS for the two  different structural 
phases at high compressions where the metal-insulator and the structural
PTs occur (denoted by a vertical arrow). The $\Delta V/V_0$
value represents a $1.(2)$\% volume collapse.
}
\end{figure}

Fig. \ref{fig:eos_lih} (left plot) presents the equation of state (EoS) 
for LiH using  
different type of  approximations to the total energy as for the 
calculation  of the lattice parameter. The Birch EoS has been used 
and the results are compared with experimental data \cite{besson}. 
The right plot shows the EoS around the B1-B2 structural transition 
and will be discussed later.
While the LDA results,  with or without the phonon contribution, are 
 unsatisfactory, the  combination of the GGA and the ZPV
leads to an excellent   
agreement with experiments up to $10\%$ of compression and to satisfactory
agreement for higher compressions.\\
\section{Excited-states properties}
\begin{figure}[h]
\twofigures[width=0.48\textwidth]{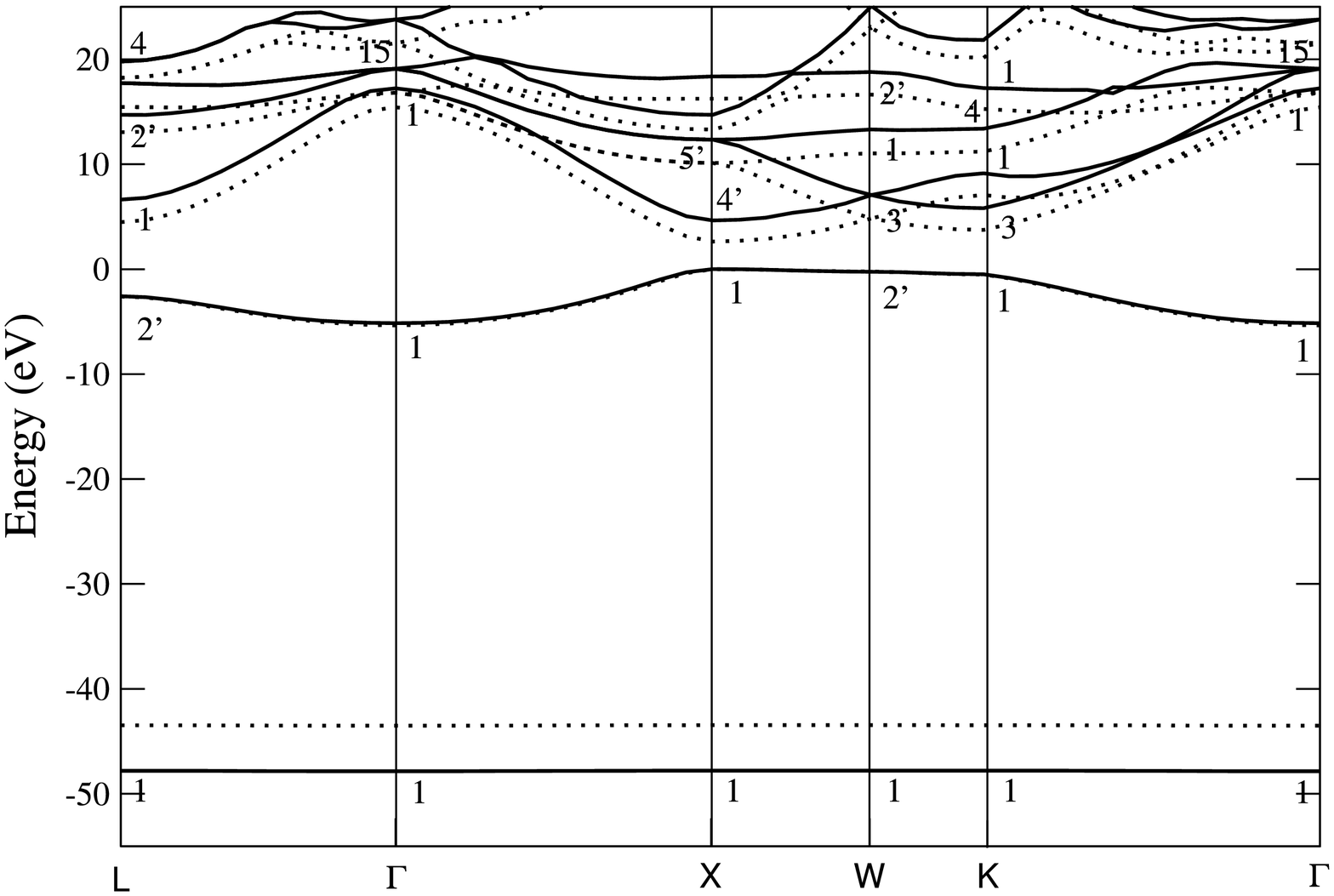}{metallisation.eps}
\caption{\label{fig:qp_lih} 
Calculated LDA (dashed lines) and GW (full lines) electronic band
 structures of LiH along some high-symmetry directions. The calculation are
performed for the B1 phase (NaCl structure) and at the equilibrium volume 
(lattice parameter $= 7.72$ a. u.).
}
\caption{\label{fig:metallisation} 
Pressure evolution of the calculated LDA (dotted lines) and GW (full lines) 
electronic excitation energies in the B1  phase. 
Circles and squares represent, respectively, the LDA and the GW calculated points. 
The dotted and full lines are used as a guide to the  eye. 
The symbol attached to each curve denotes the symmetry of states 
involved in the electronic transitions.
The LDA found, incorrectly, 
a closure of the band gap before the structural transition takes place, 
whereas the GW calculations lead to a simultaneous metal-insulator and 
structural  PTs.
}
\end{figure}

We studied the excited state properties within the framework 
 of the GW approximation to the selfenergy\cite{hedin}. 
The purpose of the GWA method is to 
 provide a rigorous formulation for the quasiparticle properties
 based on Green's function approach.
Briefly, the quasiparticle (QP) energies $E_n({\bf k})$ and 
wave function $\psi_{{\bf k}n}({\bf r})$
are determined from the solution of the QP equation
\begin{eqnarray}
(T+V_{ext}+V_{h})\psi_{{\bf k}n}({\bf r}) + \int d^3r^{\prime}
 \Sigma({\bf r},{\bf r}^{\prime},E_n({\bf k}))\psi_{{\bf k}n}({\bf r}^{\prime}) 
= E_n({\bf k})\psi_{{\bf k}n}({\bf r}) ~~~~~~~~
\label{eq:qp_psi}
\end{eqnarray}
where $T$ is  the kinetic energy operator, $V_{ext}$ the external potential 
 due to the ion cores, $V_{h}$ the average electrostatic (Hartree) potential,
 and $\Sigma$ the electron selfenergy operator, written in the GW approximation as
\begin{equation} \label{self_energy}
\Sigma({\bf r},{\bf r}^{\prime},\omega)=\frac{i}{2\pi}\int d\omega' 
G({\bf r},{\bf r}^{\prime},\omega+\omega^{\prime})e^{i\delta\omega^{\prime}}
W({\bf r},{\bf r}^{\prime},\omega^{\prime})
\end{equation}
where $G$ is the one-electron Green's function, $W$ the screened interaction 
 and $\delta$ a positive infinitesimal.
In practice, the quasiparticle energies are obtained using a 
a Taylor expansion of the selfenergy:
\begin{eqnarray}\label{quasiparticle_energy_final}
{\rm Re}E_n({\bf k}) = \epsilon_n({\bf k})+ Z_{n{\bf k}} \times
[\langle\Psi_{{\bf k}n}|
{\rm Re}\Sigma({\bf r},{\bf r}^{\prime},\epsilon_n({\bf k}))|\Psi_{{\bf k}n}\rangle
- \langle\Psi_{{\bf k}n}|V_{xc}^{LDA}(r)|\Psi_{{\bf k}n}\rangle]
\end{eqnarray}
with
\begin{equation}\label{Renormalization}
Z_{n{\bf k}}=[1-\langle\Psi_{{\bf k}n}|
\frac{\partial}{\partial\omega} \textrm{Re}\Sigma({\bf r},{\bf r}^{\prime},
\epsilon_n({\bf k}))
|\Psi_{{\bf k}n}\rangle]^{-1}.
\end{equation}
In our case, the screened Coulomb interaction $W$ is calculated in the 
random-phase approximation (RPA):
 no plasmon-pole model is invoked. More details about our implementation 
are given elsewhere\cite{lebegue}.
In Fig \ref{fig:qp_lih}, the quasiparticle band structure of LiH is
 presented. The minimum band gap is found to occur at the $X$ point.
 The LDA produced an underestimated value of $2.64$ eV, whereas
 GW approximation brings it to $4.64$ eV, within 7\% of the experimental
result\cite{baroni} of 4.99 eV.  Others transitions show 
 a considerable improvement compared  to experiment (see Table \ref{tab:gw_exp})
\footnote{In order to have well converged results, the GW calculations have been performed 
using $64$ {\bf k}-points in the full Brillouin zone and a size of 
the reciprocal-space polarisability matrix of $ 169 \times 169$ (See Ref. \cite{lebegue}).}.
In our calculation, the value of the renormalization factor Z defined in Eq. 
 (\ref{Renormalization}) of $ \sim 0.7-0.8$ for the valence top band and
 for the lowest conduction band indicates well defined quasiparticles.
A self-interaction correction calculation\cite{hama} reported a 
 minimum band gap at the equilibrium lattice parameter of $4.93$ eV and 
of 3.93 eV at $60 \%$ compression. In our case, we obtain values of $4.64$ eV
 and $3.97$ eV, respectively. It is interesting to notice  that the difference is clearly
 higher at normal pressure. However, it is difficult to reach a general
 conclusion since the comparison is biased by the use of different methods and basis sets. 
Moreover, only a simplified SIC scheme has been used in Ref. \cite{hama}.
\vspace{0.4truecm}
\begin{table}[h]
\caption{\label{tab:gw_exp}
Comparison of our LDA, COSEX, and GWA  electronic transitions and 
valence bandwidths with others available results. Data in parentheses are results 
when the denominator of the Green's function is updated with QP energies.}
\begin{center}
\begin{largetabular}{lccccc}
\hline
\hline
                           & LDA$^1$   & COHSEX$^1$ & COHSEX$^2$ & GW Approximation$^1$ & Expt.$^2$ \\
\hline
${\bf X}_{1v}$ $\rightarrow$ ${\bf X}_{4^{\prime}c}$        & 2.64 & 5.87 & 5.24 & 4.64 (4.84) & 4.99  \\
${\bf L}_{2^{\prime}v}$ $\rightarrow$ ${\bf L}_{1c}$        & 7.09 & 10.65 & 9.45 & 9.20 (9.53) & 9.0 \\
${\bf X}_{1v}$ $\rightarrow$ ${\bf X}_{5^{\prime}c}$        & 10.12 & 14.01 & 13.58 & 12.34 (12.64) & 13.5 \\
$\Delta_{v}={\bf X}_{1v}-{\bf{\Gamma}}_{1v}$       & 5.35 & 5.71 & 7.16 & 5.15 (5.48) & 6.3$\pm$1.1;6.0$\pm$1.5\\
${\bf X}^{core}_{1v}$ $\rightarrow$ ${\bf X}_{4^{\prime}c}$ & 46.10 & 54.59 & 58.82 & 52.43 (53.93) & 58.4;57.8\\
${\bf X}^{core}_{1v}$ $\rightarrow$ ${\bf X}_{5^{\prime}c}$ & 53.58 & 62.73 & 67.16 & 60.13 (61.73) & 66 \\
${\bf L}^{core}_{1v}$ $\rightarrow$ ${\bf L}_{2^{\prime}c}$ & 56.52 & 65.27 & 70.87  & 62.53 (64.20) & $>$72;70.7\\
${\bf L}^{core}_{1v}$ $\rightarrow$ ${\bf L}_{4c}$          & 58.93  & 68.59 & 73.11 & 65.56 (67.24) & $>$72;70.7\\
\hline
\hline
\newline
\end{largetabular}
\end{center}
$^1$ Present work; $^2$ Ref \cite{baroni}.
\end{table}

The overall agreement of the  previous study of the excited states of LiH 
with experiment is  fortuitous and  can be traced back to the use of 
multiple approximations\cite{baroni}. 
In particular, (1) the dielectric matrix is computed with a model function, 
(2) the selfenergy is treated in the core-hole and screened exchange (COHSEX) 
approximation\cite{hedin}, i.e.,  the dynamical
 correlation effects are neglected  and finally, (3) the Hartree-Fock
 wave functions are used as basis set. To compare satisfactorily
our calculation with this simplified one, we carried out 
a  calculation with the same COHSEX approximation  and showed that, indeed, 
 the neglect of the dynamic correlations leads to larger electronic transition 
energies than the full GW calculation. 

The most interesting point about LiH is the possibility of a pressure-induced 
MIT. Despite the fact that it is a well studied 
problem \cite{griggs,behringer,olinger,hama,hammerberg,kondo}, 
it seems that it has never been completely solved.
Our LDA calculation under pressure (see dotted lines in Fig \ref{fig:metallisation}) 
agrees  well with those of Hama {\sl et al.}\cite{hama}.
At low pressure the valence and  the lowest conduction bands are formed by 
a bonding and antibonding states of the hydrogen 1$s$ and  Lithium 2$s$ states
as shown in  Fig. \ref{fig:qp_lih}.
At high pressure the bonding anti-bonding band gap becomes larger,  and the
bottom of the lowest conduction band  of 
 lithium 2$p$ character  feels a negative pressure and moves towards lower energies, 
reducing drastically the energy band gap. The closure of the band gap finally occurs 
at the $X$ high-symmetry point of the B1 phase and the LDA MIT takes place at 
around $29\%$ of the equilibrium volume\cite{thermodynamics} 
a bit lower than the result of Ref. \cite{hama}.
Nevertheless, the LDA is well known to underestimate energy band gap and therefore to 
underestimate the volume compression of  
the MIT. This motivated us to 
apply our GW method to make a more rigorous study of this PT.
In this case (see full lines in Fig \ref{fig:metallisation}), the 
metallic transition is found to occur at $23 \%$ of the equilibrium volume, well
below our LDA value. The scenario  of the band gap closure remains the same as in the
LDA.  In term of pressure, this brings us to a value of $580$ GPa. 
The disagreement with previous reported results can be traced back to the use of 
the LDA instead of the GWA.  Hama {\sl et al.}\cite{hama} predicted
 a band gap closure at  $226$ GPa, and Kondo and Asaumi\cite{kondo} used first-order 
Murnaghan EoS to find a  metallisation pressure of $400$ GPa. 
However this electronic pressure is irrelevant, since this metallic transition competes with a 
structural PT from the B1 semiconducting phase to the B2 metallic phase at much lower
pressure.  
This structural transition was not considered by Kondo and Asaumi\cite{kondo} and by 
Hama {\sl et al.} \cite{hama} but was studied later by  Ahuja {\it et al.}\cite{ahuja}. 
In our case, we have found this structural transition at about 
 $29 \%$ of the experimental equilibrium volume;   Ahuja {\sl et al.}\cite{ahuja} 
found  it at about $33 \%$. 
At this volume, the band-gap  closure is already obtained  within the LDA, whereas 
GW still predicts an insulating state. 
As both LDA and GW calculations predict LiH to be a metal in the B2 phase,  
at the structural transition volume, the MIT is 
induced by a structural change and is of type one according to BZL classification\cite{czl}.
We have found that the PT is accompanied by a volume  collapse  of  
1.(2)\% and a band-gap closure of more than 1 eV\cite{thermodynamics}. 
This PT can be achieved experimentally  since an experimental 
pressure of the order of $332$ GPa is nowadays within reach. 
\section{Conclusion}
In this Letter, lithium hydride has been revisited in many different aspects.
We have shown that both the gradient correction and the zero-point
 motion are crucial for the correct description of the ground state of LiH.
We have also shown that the knowledge of the quasiparticle band structure under
pressure is crucial for the prediction of the MIT.
In particular, we have pointed out to the importance of applying the GW calculation
 to LiH by showing, for the first time, that the MIT happen 
simultaneously with a structural PT. The LDA predicted firstly 
the MIT and then under further compression the structural 
transformation.  We hope that our work will stimulate further experimental
studies of the MIT in LiH.
\begin{acknowledgments}
Two  of us (S. L. and W. E. P.) were 
supported by DOE grant DE-FG03-01ER45876, and 
M. A.  was supported in part by the National Science
Foundation under Grant No. PHY99-07949.
Supercomputer time was provided by the CINES (project gem1100) on the IBM SP4.
\end{acknowledgments}

\end{document}